\documentclass[12pt]{article}
\usepackage{color}
\usepackage{latexsym}
\usepackage{amsmath}
\usepackage{amsfonts}
\usepackage{color}
\usepackage{graphicx}
\usepackage{amssymb}
\usepackage{indentfirst} 
\numberwithin{equation}{section}
\newcommand{\be}{\begin{equation}}
\newcommand{\ee}{\end{equation}}

\newcommand{\mL}{{\mathcal L}}

\title{
Revisiting   the dynamics of a charged spinning body\\ in   curved spacetime} 
\author{
\\ 
\vspace*{0.3cm}
K. Andrzejewski\footnote{University of  Lodz, Faculty of Physics and Applied Informatics.  Poland, Lodz,   Pomorska 149/153,	90-236, e-mail: krzysztof.andrzejewski@uni.lodz.pl}
 \vspace*{0.3cm}
}
\date{}
\begin{document}
\maketitle 
\begin{abstract} 
  We analyse the motion of the spinning body  (in the pole-dipole approximation)  in the  gravitational and electromagnetic fields described by   the Mathisson-Papapetrou-Dixon-Souriau  equations. 
 First,     we  define  a novel spin  condition  for the  body  with the magnetic dipole moment proportional to spin,    which  generalizes     the one  proposed by  Ohashi-Kyrian-Semer\'ak for gravity.   As a result, we get the whole family of charged  spinning particle models in the 
  curved spacetime  with remarkably simple dynamics (momentum and velocity are   parallel).   Applying the reparametrization procedure, for a specific   dipole moment, we obtain equations of motion  with constant mass and gyromagnetic factor. Next, we show that these equations follow from an effective Hamiltonian formalism, previously interpreted as a  classical model of the charged Dirac particle.     
\end{abstract} 
\newpage 
\section{Introduction}
\label{s0}
The  dynamics of the  classical relativistic  spinning  bodies in  the external  fields  have  been the subject  of intense studies since the pioneering works \cite{b1a}-\cite{b1e}.     
For the    electromagnetic and  gravitational backgrounds  such investigations are  usually made  in the ``pole-dipole" approximation, i.e.  when the body is  small (the external field does not change significantly through the body)  and  does not  itself  contribute to the fields. Then,   it can only  be   
characterized by mass, spin (angular momentum) and (for charged objects)  electromagnetic dipole moments; 
all  higher multipoles are neglected. In this context, of particular interest is the description of the point-like (``elementary'') objects   exhibiting  internal  angular momentum  and its relation to   quantum spin.  
\par 
In the pole-dipole approximation the motion  can be described by body's representative worldline and the set of equations for momentum and spin tensor, the so-called Mathisson-Papapetrou-Dixon-Souriau (MPDS)  equations \cite{b2a}-\cite{b2ee}.   However,  to determine  the dynamics   the MPDS equations  have to be  supplemented by  additional conditions, the so-called  spin supplementary conditions (SSC). 
Various   SSC have been proposed and  investigated    over the years see,   among others,  \cite{b1a,b2a,b2b} and \cite{b3a}-\cite{b3o},
 as well as     \cite{b5a} for a  detailed  discussion and more  references;  despite this fact, their physical meaning  and  the selection  of the most appropriate  one     still seems to be  an  open problem. 
\par  The two most popular  and having  the  long history  are  the  Tulczyjew-Dixon (TD) \cite{b2b,b3e} and  Frenkel-Mathisson-Pirani (FMP) \cite{b1a,b1c,b3d} conditions. Both  conditions  have strengths and weaknesses identified   in various investigations.  The common  problem, however,   is  that   for both of them  the momentum and velocity  are not parallel and, consequently, the resulting equations of motion are very  complicated. Such a situation  contrasts with  some  effective approaches proposed for the   description of spinning point-like particles \cite{b10a}-\cite{b10f}. In the case of the gravitational  fields,  some light has been recently  shed on  the above problem by a  new form of  the  SSC  proposed  by Ohashi,  Kyrian, and Semer\'ak (OKS) in Refs. \cite{b3h,b3i}.  It turns out that  under the OKS condition the momentum and velocity  are parallel  (with a constant mass) and the motion fits into an effective theory of spinning particle in the  gravitational  field \cite{b4a};  this significantly simplify the study of  the dynamics.  In view of this,  the question is  whether we can extend the OKS condition  to electromagnetic fields and charged  particles  (in the curved spacetime).  Such an extension could  be useful in the analysis  of  the functional analogies between the  motion of spinning particles in gravitational and electromagnetic backgrounds, see e.g.  \cite{b6a,b6b},   or enable to find  relations with other  effective theories. Such a   question is also motivated by the double copy conjecture, which attempts to explain   gravitational processes  by means of gauge fields (for its   classical  aspects see  e.g. \cite{b7a}-\cite{b7e}) and thus  prompts a comparison of the motion of bodies  in the gravitational and electromagnetic fields, see   \cite{b8a,b8b} for spinless case. 
In this work, we will try to analyse these issues and address the above questions.   
\section{First observations}
\label{s1}
Let us  consider a classical test body  which is so small
that  multipoles beyond the dipole can be neglected. Moreover, 
taking into account the context of  the spinning   point-like (elementary) particles,  we skip the electric dipole moment and assume that  the magnetic dipole  tensor  is proportional to the spin tensor\footnote{We follow the definitions from Refs. \cite{b2b,b2c};  for the wider  discussion (and references)  concerning  the electromagnetic  dipole moments see  Ref.   \cite{b6b}.} 
$
M^{\alpha\beta}=kS^{\alpha\beta},
$
where $k$ is a function, in general  depending on $x,F,S,P$. 
Then  it turns out that  the dynamics  can be reduced to the motion of a  test particle described by the  reference worldline $x^\alpha(\tau)$  together with   four-momentum $P^\alpha$, spin tensor $S^{\alpha\beta}$ and the function $k$  satisfying along this worldline the  Mathisson-Papapetrou-Dixon-Souriau  (MPDS) equation  
\begin{align}
\label{mpds1}
\frac{D P^\alpha}{D\tau}&=q{F^\alpha}_\beta\frac{d x^\beta}{d \tau }-\frac 12{R^\alpha}_{\beta\gamma\delta}S^{\gamma\delta}\frac{dx^\beta}{d\tau}+\frac  k 2 S^{\beta\gamma}D^\alpha F_{\beta\gamma}, \\
\label{mpds2}
\frac{D S^{\alpha\beta}}{D\tau}&=P^\alpha\frac{dx^\beta}{d\tau}-P^\beta\frac{dx^\alpha}{d\tau}+k{F^\alpha}_{\gamma}S^{\gamma\beta}-kS^{\alpha\gamma}{F_{\gamma}}^{\beta},
\end{align}
where $\tau$ is the proper time  parameter;  to  simplify notation  we will write, depending on the context,   $k(\tau)$ or $k(F,S,...)$.
\newline 
Before we go further, a few remarks are in order. First, in general the momentum is not proportional (parallel) to the velocity.  Second,  we assume   that the masses
\be  \label{e1}
M^2\equiv-P_\alpha P^\alpha, \quad m\equiv -P_\alpha\frac{d x^\alpha}{d\tau},
\ee
measured in the zero three-momentum and in the zero three-velocity frames,  respectively,  are positive. This is not guaranteed by the MPDS equation and the breakdown of this assumption  suggests that  the pole-dipole approximation may be not valid.
\par
Now,   let us make an observation that will  be useful.   Namely, for the  constant-sign   function $k$ (to fix attention, positive)   there is a change of the parametrization  such that  for  the new parametrization  the factor $k$  in the MPDS equations  is constant, i.e.   $k=k_0>0$\footnote{We do not specify $k_0$; however, to make contact with the  non-relativistic limit we can take  $k_0=q/m$.  }. In fact, defining the new parameter $\tilde\tau$  as follows 
 \be  \label{e2}
 \tilde\tau(\tau)=\frac{1}{k_0}\int^\tau  k,
 \ee
 we find  that $\frac{d\tilde\tau}{d\tau}=k/k_0>0$; thus  there exists an inverse function $\tau=\tau(\tilde\tau)$. Now,  we easily check that such a  change of the  parametrization  leads to the  constant factor $k_0$ (instead of $k$) in the MPDS equations. Moreover, in the new  parametrization the  length of the  velocity vector  reads   
 \be  \label{e3}
 -\frac{dx^\alpha(\tilde\tau)}{d\tilde\tau} \frac{dx_\alpha(\tilde\tau)}{d\tilde\tau}
 \equiv
 -\frac{dx^\alpha(\tau(\tilde\tau))}{d\tilde\tau} \frac{dx_\alpha(\tau(\tilde\tau))}{d\tilde\tau}=\frac{k_0^2}{k^2(\tau(\tilde\tau))}
 \equiv\frac{k_0^2}{k^2(\tilde\tau)}.
 \ee
 \par 
  The MPDS equations  are not sufficient to   find   the dynamics.  To close the system we  add three   constraints,  the so-called spin supplementary condition (SSC), of the form 
\be
\label{ssc}
	S^{\alpha\beta} V_\alpha=0,
\ee
where $V^\alpha$   is a vector field  defined, at least, along $x^\alpha(\tau)$. This condition can be interpreted as  the choice of   worldline $x^\alpha(\tau)$ such that 
 the center of mass is measured by some observer moving with  the velocity $V^\alpha$. 
Thus, usually 
$V^\alpha$ is assumed to be time-like (without lost of generality $V^\alpha V_\alpha=-1$); however, in our considerations we will admit   also the  null-like case  $V^\alpha$, $V^\alpha V_\alpha=0$, see Sec. \ref{s2}.  
\par 
The  choice $V^\alpha =\frac{dx^\alpha}{d\tau}$ has been proposed  by Frenkel for  the electromagnetic field  \cite{b1a}  and later for gravity by Mathisson-Pirani \cite{b1c,b3d}; 
however,  such a condition  leads  to some problems  with the  uniqueness of solutions. Namely,  even in the flat spacetime it does not yield the unique worldline (it depends on the choice of initial conditions); though,    it has been recently  argued that this  ambiguity is  not physically relevant, as it provides   alternative descriptions of the  body motion \cite{b5a,b5aa}.  
Another SSC has been proposed by Tulczyjew and Dixon; namely, they considered  $V^\alpha=P^\alpha/M$. Then,  the dynamics is  unambiguous; however,  some arguments have been made that the 
canonical momentum of massless spinning particles does not have to  be time-like \cite{b5aaa} or  in the ultra-relativistic limit (when the particle velocity approaches to the speed of light)  the acceleration in the direction of velocity grows up to infinity 
\cite{bx1,bx2}. Putting aside the above problems, let us stress that  for the above spin  conditions  the velocity-momentum relation is very complicated, what makes   the  analytical   considerations very difficult. This   is especially   evident in the presence of the  electromagnetic field; even if  the electromagnetic dipole moment equals zero, see e.g. \cite{b3f,b5b}.
\par  
Obviously,  other   SSC have been also   proposed, see e.g.  \cite{b3a,b3b,b3c,b3g},   but  usually they   refer to the form of background  fields, thus they  are called  the background conditions. Such conditions are matched to a  special form of the external  fields, consequently they can simplify equations of motion in these fields.   One  such  example  is   $V=\partial_t$  for the Schwarzschild metric considered in Ref. \cite{b3c} (the CP condition), another example is   related to the  Vaidya  metric \cite{b3g}.  
\par 
For the gravitational background  ($F_{\alpha\beta}=0$), an alternative approach has been  recently proposed by the   Ohashi,  Kyrian, and Semer\'ak  \cite{b3h,b3i}. Namely, they postulated  $S^{\alpha\beta}V_\beta=0$ where $V^\alpha$ is a vector field parallelly transported  along $x^\alpha(\tau)$, i.e. $DV^\alpha/D\tau=0$. Then, it turns out that  the momentum $P^\alpha$ is proportional to the velocity, $P^\alpha =mdx^\alpha/d\tau$, and $M=m$ is constant. This, in turn, significantly  simplifies the   MPDS  equations for the gravitational background; in particular, the spin tensor is parallel-transported,  
$
 \frac{D S^{\alpha\beta}}{D\tau}=0.
$
 Such a  simplification   allows for   a  more analytical considerations.    
Moreover,  it has been  recently   realized that the   OKS approach is related  to   an effective Hamiltonian formalism for  spinning point-like objects \cite{b4a}.     
In view of the above   the  OKS condition has several advantages.  We will show that it can be naturally   extended to  electromagnetic  fields (in the curved spacetime).  
\section{Extension of  OKS condition to  electromagnetic fields}
\label{s2}
For the MPDS equations  \eqref{mpds1} and \eqref{mpds2} let us define the  spin condition of the form  ${S^\alpha}_\beta V^\beta=0$, where $V^\alpha$ is a time-like or null-like vector field along $x^\alpha(\tau)$ such that 
\be
\label{ec}
\frac{DV^\alpha}{D\tau}=k{F^\alpha}_\beta V^\beta.
\ee 
Obviously, for $F_{\alpha\beta}=0$ and time-like $V^\alpha$ we recover the OKS condition   ($V^\alpha$ is parallel-transported).  Since $F_{\alpha\beta}$ is a skew-symmetric matrix the lenght of $V^\alpha$ is constant, thus we can assume that $V^\alpha V_\alpha= -1$ or $V^\alpha V_\alpha= 0$. In Sec. \ref{s1}  we have seen that there is a parametrization $\tilde\tau$ such the MPDS equations can be transformed into the ones  with  $k$ constant. Applying  this reparametrization to the condition \eqref{e3}  we arrive, in agreement  with this observation, at eq.  \eqref{ec} with $k=k_0=const$. Thus the condition \eqref{ec}  is reparametrization compatible with the MPDS equations (the change of the  parametrization results in the same both for the MPDS equations and  \eqref{ec}). 
\par We begin the   analysis of the condition \eqref{ec} by showing that, in  the presence of  the electromagnetic field, it  also leads to  the same crucial property as the  OKS condition  for gravity, i.e. the momentum and velocity are parallel. In fact,  contracting  eq. \eqref{mpds2} with $V_\beta$ and using the condition \eqref{ssc} as well as \eqref{ec} we arrive at the identity 
\be  \label{e5}
P^\alpha(V_\beta\frac{dx^\beta}{d\tau})-(P^\beta V_\beta)\frac{dx^\alpha}{d\tau}=0;
\ee  
 taking into account that $d x^\alpha/d\tau $ is time-like and $V^\alpha$ is time-like or null-like we get desired proportionality.
 Now, for our further  purposes, let us take  an arbitrary parametrization $\lambda$. Then, for  $U^\alpha=\frac{dx^\alpha}{d\lambda}$  we  have
 \be  \label{e6}
-U^\alpha U_\alpha =f(\lambda ), \quad f>0;
 \ee 
 in particular,  for proper time ($\lambda=\tau$) we get $f=1$, while for $\lambda=\tilde\tau$ given by eq. \eqref{e3} we get $f(\tilde \tau)=k_0^2/k^2(\tilde\tau)$. Now,   eq. \eqref{e5} can be easily rewritten in the new parametrization $\lambda$ and next  multiplying it by $\frac{dx^\alpha}{d\lambda}$ we arrive at the identity 
 \be
 \frac{m}{f}=\frac{P_\beta V^\beta}{V_\beta\frac{dx^\beta}{d\lambda}}.
 \ee
 This together with eq. \eqref{e5} (in terms of the new parametrization)  yield 
\be  \label{e11}
P^\alpha=\frac{m}{f}U^\alpha, \quad fM^2=m^2.
\ee
 In particular,  for the proper time parametrization  ($f=1$) we  have $P^\alpha=m\frac{dx^\alpha}{d\tau}$ and both masses coincide; in general, the mass $m$  depends on the parametrization (in contrast to $M$). Moreover,   let us stress  that,  under condition (3.1) (in the presence of  the electromagnetic field),       mass does not have  to  be a constant; in contrast to   the OKS condition  for  gravity alone.  Finally, by virtue of  eq. \eqref{e11} the second part of the MPDS equations \eqref{mpds2} takes the form
\be  \label{e12}
\frac{D S^{\alpha\beta}}{D\lambda }=k{F^\alpha}_{\gamma}S^{\gamma\beta}-kS^{\alpha\gamma}{F_{\gamma}}^{\beta}.
\ee
\par
Up to now we do not specify $k$ and we do not use the first part of the MPDS equation,  see  eq. \eqref{mpds1}. Differentiating  the second equation in  \eqref{e11}  and  using eq.  \eqref{mpds1} we obtain  the following relation 
\be  \label{e13}
\frac{d m}{d\lambda }=\frac{m}{2f}\frac{d f}{d\lambda}-\frac{1}{2}kU^\alpha S^{\beta\gamma} D_\alpha F_{\beta\gamma};
\ee
where  $k=k(\lambda)$ denotes the  factor in the new parametrization,  i.e. $\frac{d\tau}{d\lambda}k(\tau(\lambda))$.
Using    eq. \eqref{e12} it can be rewritten in the form
\be  \label{e14}
\frac{d m}{d\lambda }=\frac{m}{2f}\frac{d f}{d\lambda}-\frac{1}{2}k \frac{d}{d\lambda} (S^{\beta\gamma}F_{\beta\gamma}).
\ee
To analyse this condition let us take proper time parametrization $\lambda=\tau$,  equivalently $f=1$.  Then, for  $k$ being  a function of $a=F_{\alpha\beta}S^{\alpha\beta}$ only,  more precisely $k(a)=h'(a)$  where $h$ is a  function of one variable,  eq. \eqref{e14} can be integrated  explicitly yielding  the following direct form of the mass parameter   
\be  \label{e16}
m=m_0-\frac 12 h(F_{\alpha\beta}S^{\alpha\beta});
\ee
where $m_0$, for the choice $h(0)=0$, can be identified with the ``bare" mass, i.e. for $F=0$. 
\par In summary, for the proper time parametrization and a function $h$,  the spin condition \eqref{ssc} with $V^\alpha$  satisfying \eqref{ec} reduces the second part of the MPDS equations  to eq. \eqref{e12} (with $\lambda=\tau$) and   the first one takes  the form
\be  \label{e17}
\frac{D }{D\tau}(m\frac{dx^\alpha}{d\tau})=q{F^\alpha}_\beta\frac{d x^\beta}{d \tau }-\frac 12{R^\alpha}_{\beta\gamma\delta}S^{\gamma\delta}\frac{dx^\beta}{d\tau}+\frac  1 2 k S^{\beta\gamma}D^\alpha F_{\beta\gamma},
\ee   
where 
\be 
\label{e18b}
k=h'(a)= h'(F_{\gamma\delta}S^{\gamma\delta}),
\ee
and $m$ is given by \eqref{e16}.  
\par 
Now, let us make a few comments.  
First, we see that eqs. \eqref{e12} and  \eqref{e17} contain    $x^\alpha$ and $S^{\alpha\beta}$ only,  they  do not contain $V^\alpha$. This is closely related to definition  \eqref{ec} of $V^\alpha$. In fact, $V^\alpha$ can be  reconstructed (along $x^\alpha(\tau)$) from  the initial condition  $V^\alpha(\tau_0)=0$      by means  of  eq.  \eqref{ec}   (we have a linear set of differential equations for $V^\alpha$ and thus it possesses a  global solution). In consequence, the choice of $V^\alpha$ (and thus various forms of $S^{\alpha\beta}$ and $x^\alpha$) are  encoded in the initial conditions $S^{\alpha\beta}(\tau_0)V_\beta(\tau_0)=0$; $V^\alpha(\tau_0)$ is an arbitrary (normalized) vector. For example, we can put  $V^\alpha(\tau_0)=\frac{dx^\alpha}{d\tau}(\tau_0)$  then our condition  agrees with the FMP condition  at the initial time; however,  for further times the two approaches  are different.  
\par
Second, in addition to  the  original OKS approach the above procedure  holds  also for    a null-like vector ($V^\alpha V_\alpha=0$).  Since the   multipole scheme for the extended body (in particular the center of mass) is  associated with a  time-like vector (corresponding to  the velocity of some observer),   such a possibility  seems  speculative from the physical point of view. However, let us note that  the null-like case can be treated as    an idealization  of the ultra-relativistic  limit (the limiting  centroids \cite{b3i} are defined by the systems which move almost at the speed of light; in other words, they define  the minimal worldtube of a spinning body). Given the above observation, let us note that for gravity ($q=k=0$) there is  a natural   place  where  such  null-like vectors can be used.    Namely,  let us consider pp-waves metric (in particular the   plane gravitational waves)
\be
\label{ppw}
g=K(x^1,x^2,u)du^2+2dudv+(dx^1)^2+(dx^2)^2.
\ee 
 For such spacetimes  there exists the   null-like  Killing  vector    $V=\partial_v$ which can be used as the SSC. The  motivation behind this choice is that   the  $u$-coordinate  can be considered as a  ``substitute" of  time \cite{b3ii}.   Thus,  such a choice corresponds to the  CP condition (in the  light-cone coordinates).  Then, one  can  easily check that $V$  satisfies  the OKS condition. In consequence,    the spin is parallel-transported.  Moreover, for our spin condition we have that   $S^{u\alpha}=0$; this together with the form of the metric \eqref{ppw}, after straightforward computations, yield that  the   term ${R^\alpha}_{\beta\gamma\delta}S^{\gamma\delta}\frac{dx^\beta}{d\tau}$ in eq. \eqref{mpds1} is equal to zero, thereby the motion is geodesic.  In this  way  we extend   the results  of Ref. \cite{b9} to the pp-waves metric and    included them into   the OKS approach. 
\par
 Third,  the pole-dipole approximation  applies to   the motion of a extended body; however, from the very beginning  some attempts have been made to interpret it for point-like  objects as well.  Of course, such attempts immediately lead  to   a problem; namely,    for the  extended body we are  free  to  define a representative  point by which we want  to describe the  motion --  in the case of  the point particle we  have  no such freedom.  Moreover, due to  M\o ller's reasoning \cite{b3iii} a classical spinning body must have a finite extension, defined  by  width of the
 worldtube of centroids.  In consequence,  it  is no longer evident which spin supplementary condition should be adopted to describe the point-like particle.   On the other hand,  there are a number of  problems which motivate to   study   relativistic spinning particles; for example, classical limit of the Dirac particle,  non-minimal spin-gravity coupling, higher spins theories; in general, semi-classical description of elementary particles. As a result,  some  alternative approaches and effective theories were proposed, see among others \cite{b2e} and  \cite{b10a}-\cite{b10f};  they define relativistic  spinning  point-like objects by assigning them  an overall position, momentum and spin.    From  this  point of view, with each function $h$ we can associate a model  of an elementary classical  particle  defined by  $k=h'$  and  the set of  equations  \eqref{e12},  \eqref{e16}, \eqref{e17}, together with the initial condition for $V^\alpha$.  
 Moreover, by    virtue of eqs. \eqref{e11} (with $f=1$)  and \eqref{e16}, we  have the following  equation of state  which  describes  our model 
 \be  \label{e18}
 M^2\equiv - P^\alpha P_\alpha =(m_0-\frac 12 h(F_{\alpha\beta}S^{\alpha\beta}))^2.
 \ee  
Such a model, due to the fact that momentum and velocity  are parallel as well as $m=M$,  is much  simpler than the one obtained by  means of the  TD condition, see Refs. \cite{b2c,b2e}. 
\par
Another aspect, related to fact that  momentum is proportional to  velocity, concerns the so-called hidden momentum. At the dipole order this is especially interesting for the electromagnetic interaction, since    there may  be a part  of momentum   of  mechanical nature, see e.g. \cite{bb1,bb2}.   Let us analyse this issue in more detail.  
To this end let us consider an arbitrary spin supplementary condition\footnote {Since the issue concerns  electromagnetic field,   we   restrict ourselves here to the  the Minkowski spacetime; dot denotes the ordinary derivative with respect  to proper time.} $S^{\alpha\beta}V_\beta=0$, then multiplying the second part of the MPDS  equations by $V^\alpha$   and next using  again the SSC we arrive at the equation 
\be
P^\alpha=\frac{1}{\dot x^\beta V_\beta}\left( (P_\beta V^\beta)\dot x^\alpha - S^{\alpha\beta}\dot V_\beta+kS^{\alpha\gamma}{F_\gamma}^\beta V_\beta\right). 
\ee  
Thus the momentum is  the  sum of three components 
\be
P^\alpha= P^\alpha_{kin}+P^\alpha_{hidI}+P^\alpha_{hidD};
\ee 
where,  accordingly to  the terminology of Ref. \cite{b5a},  the ``kinetic" part $P_{kin}^\alpha=(P_\beta V^\beta)\dot x^\alpha/(\dot x^\beta V_\beta)$ is  related to the motion of the  center of mass, and the  so-called ``hidden" momentum consists  of two parts:  the  ``inertial"  one  $P_{hidI}^\alpha=-S^{\alpha\beta}\dot V_\beta/(\dot x^\beta V_\beta)$, and the ``dynamical" one $P_{hidD}^\alpha=kS^{\alpha\gamma}{F_\gamma}^\beta V_\beta/(\dot x^\beta V_\beta)$. The latter  $P_{hidD}^\alpha$  may contain  a  purely mechanical part  and cannot be made zero   by chaining the center of mass.   Apart from the momentum  $P^\alpha$ we can also  have  the   field momentum $P^\alpha_{em}$  directly  related to the electromagnetic fields.  Then due to the conservation law ${T^{\alpha\beta}}_{,\beta}=0 $,   the whole momentum  	$P^\alpha+P^\alpha_{em}$ should be constant (e.g.  zero for a  stationary body).
Taking various $V^\alpha$  (thus centroids)  $P^\alpha $ should be the same, but  can be made  of different  parts (depending on $V^\alpha$).    For some choices of $V^\alpha $  (see below)  $\vec P_{kin}$ can be zero, for another  choices the part  $\vec P_{hidI}$ can be zero, but (in general) not $\vec P_{hidD}$.
Now, the key point is that for $V^\alpha$ satisfying  \eqref{ec} we have  that  
\be 
\label{eid}
P_{hidI}^\alpha=-P_{hidD}^\alpha;
\ee   i.e.   for our  choice of the  centroid (defined by $V^\alpha$)  the  inertial momentum  has a very special form,  namely such that $\vec P_{hidI}+\vec P_{hidD}=0$. Then, however, there  remains   $P_{kin}^\alpha$ only,   in consequence  $P^\alpha$ is proportional  to the velocity of the centroid  ($P^\alpha=m \dot x^\alpha)$; at the same time,    the part $P_{kin}^\alpha$  takes  such a  form  that  the conservation law holds. 
\par 
To illustrate the above issue    let us  consider in the Minkowski spacetime a body with no net charge, but possessing a magnetic dipole moment $\mu^{\alpha\beta} = kS^{\alpha\beta}$, placed in the constant  electric field.  Then, eq. \eqref{mpds1} gives $\dot P^\alpha=0$.    
First,  let us take the FMP SSC. More precisely, we put $V^\alpha_0=\dot x^\alpha=(1,0,0,0)$. For such a choice of SSC, $S^{\alpha\beta}$ is constant and  the spatial momentum, $\vec P_{kin}$   and $\vec P_{hidI}$, vanish; in consequence, $\vec P=\vec P_{hidD}=k\vec S\times\vec E$, where $S^l={\epsilon_{ij}}^lS^{ij}/2$ is  the spin associated with this choice of the centroid. On the other hand,  the momentum of    the electromagnetic
field for  our system is of the form  $\vec P_{em}=-\vec \mu \times \vec E=-k\vec S\times \vec E$,  in consequence, the total moment vanishes $\vec P_{em}+\vec P=0$, as expected.  A similar situation holds for the TD condition. 
Now, let  us take  the  SSC  given by eq. \eqref{ec}.  Then as we pointed out above  $\vec P_{hidI}+\vec P_{hidD}=0$ and thus $\vec P=\vec P_{kin}$; since the momentum should be the same we have  the following  motion of the  centroid  $\dot {\vec x}=\frac {k}{m}\vec S\times \vec  E$ which ensures  that the conservation law holds.
\par
Alternatively,  for our model $P^\alpha$ is constant, thus  we can choose  the frame such that  $P'=(M,\vec 0) $. Then,  the TD condition  implies   ${S'}^{\alpha0}=0$ and ${S'}^{ij}=const.$ In consequence, the centroid moves with  the velocity $\dot {\vec x}'=\frac kM S'^{ij}E_j'$ to balance $\vec P_{hidD}'$  -- the situation is similar to the one  observed above  for  the SSC (3.1) (here,  the  kinetic term plays the same role as  the inertial term above). Next, let us analyse the spin  condition with $V^\alpha$ satisfying  \eqref{ec}.  Here  again,  $\vec P'_{hidI} +\vec P_{hidD}'=0$ as well as  $\vec P'=0$ thus $\dot {\vec x}'=0$  (i.e. $\vec P_{kin}'=0$) the centroid is at rest in this frame (though  $\vec P_{hidD}'$ is not zero, but compensated by $\vec P_{hidI}'$).
In summary, imposing various SSC or taking different  frames the  momentum $\vec P_{kin}$, $\vec P_{hidI}$ can be zero or $\vec P_{hidD}$   can  be ``hidden" by $\vec P_{hidI}$  or by $\vec P_{kin}$.
\par
It follows from the above considerations that  the centroid motion  under the condition \eqref{ec} can be quite different from that  for the FMP or TD condition (e.g. it can be described by a  straight line   instead of the point).   This, in turn, calls for a discussion on the physical meaning   and applicability range of   this spin condition.  Below,   using  the above example, we begin  such an analysis; however, a more complete discussion   would be left for further investigations, see Sec. \ref{s5}.   So let us take the pure magnetic dipole  in  the constant electric field  $\vec E=(0,0,E)$. 
Then the   (hidden) momentum 
$ \vec P=k \vec S\times \vec E$ reads  
\be 
\label{rr1}
P^a=\omega S^{3a}, \quad   P^3=0,   \quad \omega=kE;
\ee
where $a=1,2$.     
To compare the FMP and   \eqref{ec} conditions  we assume that  they  coincide  at the initial time, i.e.   $V^\mu(0)=V^\mu_0$.  Then we readily obtain that  
\be
\label{rr2}
V^\mu=(\cosh(\omega\tau),0,0,\sinh(\omega\tau)), 
\ee        
i.e.  $V^\alpha$  describes the velocity of some accelerated observer.   
The  shift  between both centroids is  given by the formula (10) from Ref. \cite{b5a}:
\be
\label{rr3}
\Delta x^\alpha=-	\frac{S^{\alpha\beta}V_\beta}{m(V)}=	\frac{S^{\alpha\beta}V_\beta}{P_\gamma V^\gamma}.
\ee 
Inserting  \eqref{rr2} into equation  \eqref{rr3} we get
\be
\label{rr4}
\Delta x^0=0, \quad  \Delta x^3=0, \quad \Delta x^a=\frac{S^{3a}}{P^0} \tanh (\omega\tau), \quad a=1,2. 
\ee
Since $V^\mu\Delta x_\mu =0$, we have that $\Delta \vec x$  is perpendicular to the velocity of the observer described by $V^\mu$  and thus $\Delta \vec x$ is the same  for the  stationary observer defined by the FMP condition. In view of eq. \eqref{rr4} we have
\be
||\Delta \vec x||=\frac{|\tanh(\omega\tau)|(S_1^2+S_2^2) }{P^0}\leq \frac{||\vec S||}{P^0}\leq \frac{||\vec S||}{M}.
\ee 
Thus   in agreement with M\o ller's  result, that the set of all shift vectors corresponding to all possible observers spans a disk of the radius $R=S/M$, we have  that the separation  between both centroids is contained in the worldtube of the  body.
\par
 On the other hand,    for   the field satisfying eq.   \eqref{ec}  the  centroid  motion  is described by the equation 
\be
\label{rr6}
\dot x^\mu=\frac{P^\mu}{M}, \quad M=m=const.
\ee 
To analyse this  issue  let us observe that,  by virtue of eqs. 
\eqref{rr1}, \eqref{rr4}, and \eqref{rr6}    the  following identity 
\be
\label{rr7}
M \dot {\vec x}=\tanh^2(\omega\tau)\vec P+P^0(\Delta \vec x)^\cdot\ 
\ee 
holds. Alternatively, after introducing the new parameter $\tau'=\tau-\tanh(\omega\tau)/\omega$,  we have that 
\be
M {\vec { x}}'=\vec P+P^0(\Delta \vec x)'\ . 
\ee
In consequence,  the transversal part of the centroid velocity (equivalently,  the kinetic momenta $\vec P_{kin}=m\dot {\vec x}$) is not equal to the velocity of the shift;  the  difference between them is caused by   the hidden momentum of the body.   
This situation  resembles the one discussed  in Ref. \cite{b5aa}. However, there the  body under consideration was free (closed system and the TD condition) and thus the hidden momentum comes from the acceleration of the observer (the inertial hidden momentum). In our case,  apart from   the accelerating observer  we have a non-closed system (i.e. beside the body we have an external  electric  field);  in consequence,   there is  a  dynamical  hidden momentum. 
To analyse this issue let us compute the hidden momentum for the centroid defined by the SSC  with $V^\alpha$ given by eq. \eqref{rr2}. To this end, first,  we find the spin components $S_*^{\mu\nu}$ satisfying the MPDS equation and $S_*^{\mu\nu}(0)=S^{\mu\nu}$. After straightforward  calculations we find that 
\be 
\label{rr9}
S_*^{03}=0,\quad S^{0a}_*=S^{3a}\sinh(\omega\tau), \quad S^{3a}_*=S^{3a}\cosh(\omega\tau), \quad S_*^{21}=S^{12}=const.
\ee
Next, using eqs. \eqref{rr9}  we get 
\be
\vec P_{hidD}=-\frac{m\vec P}{P_\alpha V^\alpha} =\frac{m\vec P}{P^0\cosh(\omega\tau)}.
\ee
Analogously or using  eq. \eqref{eid}  we can find $\vec{P}_{hidI}$.  In consequence  eq. \eqref{rr7} takes the form 
\be 
\frac {M}{P^0} {\dot {\vec { x}}}=\frac{-\sinh^2(\omega\tau)}{M\cosh(\omega\tau)}\vec P_{hidI}+(\Delta \vec x)^\cdot 
\ee
After suitable  reparametrization  $\tau'=-(\sinh(\omega \tau)-\arctan(\sinh(\omega \tau)))/\omega$ we arrive at the formula
\be
\frac {M}{P^0} {\vec { x}}'=\frac{\vec P_{hidI}}{M}+(\Delta \vec x)'\ .
\ee 
In summary, in our case $V^\alpha$ describes an accelerated observer (it varies along  the trajectory)  this in turn leads to non-trivial  motion of the centroid; however, under our condition $\vec P_{hidD}=- \vec P_{hidI}$, thus  the dynamical hidden momentum gives an additional impact, while for the FMP condition it  makes that the momentum is not proportional to the velocity of the centroid.  
\par
At the end of this section, let us note that, despite of  the very simple momentum-velocity relation, implying by the condition \eqref{ec},   the dynamics  of  mass is nontrivial even for the Minkowski spacetime and $k$ constant, say $k=k_0$  ($h(a)=k_0a=k_0 F_{\alpha\beta}S^{\alpha\beta}$).  This   makes, at the dipole order,    the electromagnetic considerations  for spinning particles more complicated than the gravitational ones (even with the OKS condition).
In addition, let us recall that  there are  effective theories  of charged spinning point-like particles, see \cite{b10a,b10b,b10e}; in contrast,  in these  theories,   the constant mass is built   in from the  very beginning.  Our next step is to show that some of these models  fit into our results exactly (and thus the   MPDS equations); the key to do   this is the reparametrization described in Sec. \ref{s1}.       
\section{Effective theories  and MPDS equations}
\label{s3}
In the previous section we showed that the SSC \eqref{ssc} with  \eqref{ec}  leads to  the   MPDS equations describing  a relatively simple model; however,  the  mass is not constant and we can choose various functions $k$'s.  So the question  arises   whether there are any natural and useful  forms of $k$. This question is also  motivated by some  effective Hamiltonian description of the spinning particles, where the mass and $k$  are constant from the very beginning.
Such theories  form  another  approach to spinning bodies, yielding  equations of motion for the point-like particles exhibiting an overall  position, momentum and spin.  They are usually obtained in two ways: starting with  a  Lagrangian
formulation  (see e.g. \cite{b3f,b3j,b11a,b11b,b11c} and references therein) or  directly by a symplectic  (Hamiltonian) description \cite{b10a,b10b,b10c,b10d,b10e,b10f,b4a,b10g,b10h}.         In the latter case the starting point is an   effective Hamiltonian formalism in the extended (due to spinning degrees of freedom)  phase space. More precisely, in presence of the  gravitational and electromagnetic fields  the phase space structure is defined by the following Poisson bracket \cite{b10a,b10b}
 \be  \label{ep1}
\{x^\alpha,P_\beta\}=\delta^{\alpha}_{\beta}, \quad \{P_\alpha,P_\beta\}=-\frac{1}{2}R_{\alpha\beta\gamma\lambda}S^{\gamma\lambda} +qF_{\alpha\beta},\quad \{S^{\alpha\beta},P_\gamma\}=\Gamma^\alpha_{\gamma\delta}S^{\beta\delta}- \Gamma^\beta_{\gamma\delta}S^{\alpha\delta},
\ee
 \be  \label{ep2}
\{S^{\alpha\beta},S^{\gamma\delta}\}=g^{\alpha\gamma}S^{\beta\delta}-g^{\alpha\delta} S^{\beta\gamma}+g^{\beta\delta}S^{\alpha\gamma}-g^{\beta\gamma}S^{\alpha\delta}.
\ee
Next,   effective Hamiltonians (satisfying some   natural assumptions, e.g. covariant and  quadratic in   momenta) are postulated. The  simplest   Hamiltonian  seems the one  discussed  in Refs.  \cite{b10a}-\cite{b10f}  
 \be  \label{eh}
H=\frac{1}{2m_0}g^{\alpha\beta} P_{\alpha}P_{\beta}-\frac{k_0}{2}F_{\alpha\beta}S^{\alpha\beta}.
\ee
\par 
Recently,  it  has been  shown, see Ref. \cite{b4a}, that  for the gravitational sector only (i.e. $F=0$) the dynamics obtained by means of \eqref{ep1}, \eqref{ep2} and  \eqref{eh} is  exactly  the same as for  the OKS condition. Now, let us consider  also  the electromagnetic backgrounds. In this case the effective equations of motion (governed by the Hamiltonian \eqref{eh})  have the  similar  form to eqs. \eqref{e12} and \eqref{e17},  however,  with constant $k$ and $m$. On the other hand, from eq. \eqref{e16} we have that  in presence of electromagnetic field $m$ is  constant   for    $h=0$ only; this implies $k=0$, which  in turn   does not  coincide with the effective approach.  Thus, for the electromagnetic fields, the constant  mass  and $k$ in the effective approach are somewhat puzzling (more generally,  there is a question about  the relation between  the MPDS equations and effective theories).  To solve this  problem  we use the reparametrization procedure discussed  in Sec.  \ref{s1}. Namely,   first we   transform the  MPDS equations, by means of the   parametrization  $\lambda=\tilde \tau$,  into  the ones with constant $k$.  Then,  the second part  of the MPDS equations (cf. eq. \eqref{e12})  takes the desired form while the first one (see eq.  \eqref{e17}) reads  
\be  
\label{e18a}
\frac{D }{D\tilde\tau}(\tilde m \frac{dx^\alpha}{d\tilde\tau})=q{F^\alpha}_\beta\frac{d x^\beta}{d \tilde \tau }-\frac 12{R^\alpha}_{\beta\gamma\delta}S^{\gamma\delta}\frac{dx^\beta}{d\tilde \tau}+\frac  1 2 k_0 S^{\beta\gamma}D^\alpha F_{\beta\gamma},
\ee
where we have  introduced the    ``effective mass" $\tilde m(\tilde \tau)\equiv m(\tilde\tau)/f(\tilde \tau)$.
Now, the question is whether  we can find a function $f$ (equivalently $k$, see eq. \eqref{e3}) such that the effective mass $\tilde m(\tilde \tau)$ is constant. By virtue of eq. \eqref{e14},  we have that 
 \be  \label{e19}
\frac{d \tilde m}{d\tilde\tau}=-\frac{\tilde m}{2f}\frac{d f}{d\tilde \tau }-\frac{k_0}{2f}\frac{d( S^{\alpha\beta}F_{\alpha\beta})}{d\tilde \tau }.
\ee
So $\tilde m (\tilde \tau)=\tilde m_0$ is constant provided
 \be  \label{e20}
f=1-\frac{k_0}{\tilde m_0}S_{\alpha\beta}F^{\alpha\beta}.
\ee
In this case, 
\be  \label{e22}
m(\tilde \tau )=\tilde m_0-k_0 S_{\alpha\beta}F^{\alpha\beta}.
\ee  
Since  the  parametrization $\tilde \tau$ is directly related to the  choice of  the function $k$ we conclude, by  virtue of eqs.   \eqref{e3} and \eqref{e6} (with $\lambda=\tilde \tau$),    that  eq. \eqref{e20} leads to  the function $k$ of the form  
\be  \label{e21}
k(a)=k(S_{\alpha\beta}F^{\alpha^\beta})=\frac{k_0}{\sqrt{1-\frac{k_0S_{\alpha\beta}F^{\alpha\beta}}{\tilde m_0}}},
\ee
(dependence $k=k(a)$ is the same  in both the  parametrizations).
\par To identify the  effective mass $\tilde m_0$ we
 compare  the momenta in the $\tau$ and $\tilde \tau$ parameters. First, we find the relation between masses in  both   parametrizations     
\be
m(\tau)=\frac{m(\tilde\tau (\tau))k(\tau)}{k_0};
\ee 
thus in the  special  case  \eqref{e22} we get  
\be
\label{e21b}
m(\tau)=\tilde m_0 \sqrt {1-\frac{k_0 F_{\alpha\beta}S^{\alpha\beta} }{m_0}  }.
\ee
Now, putting $F=0$ in the above equation, we find that both   bare masses coincide $\tilde m_0=m_0$. In consequence, $h(FS)= 2m_0-2m_0\sqrt{1-\frac{k_0FS}{m_0}}$.  
Finally,  we can  directly  confirm  our results,    the  reparametrization \eqref{e2}   transforms  eqs. \eqref{e17}  with \eqref{e21} and \eqref{e21b} into  eq. \eqref{e18a} with constant  mass $\tilde m=\tilde m_0 $ and factor $k=k_0$.
\par  By  restoring explicitly the speed of light we see that  $k$ given by \eqref{e21} is  well-defined for sufficiently weak  field; otherwise, however, we cannot apply relativistic  and  classical prescription (since  quantum effects may  come into play) or pole-dipole approximation can be no longer valid. Moreover, then   the  equation of state  reads
 \be  \label{e23}
 M^2\equiv - P^2=m_0^2-m_0k_0S^{\alpha\beta}F_{\alpha\beta},
\ee
and  it  agrees with the one obtained by means of  the  Hamiltonian  \eqref{eh} (the constant of motion) provided that  $H=-m_0/2$. Eq. \eqref{e23} can be interpreted as a classical  model of the charged   Dirac particle, see e.g.  Ref. \cite{b10b}.
Let us further note that, by means of the  TD SSC we can get analogues   conclusions, see 
 Refs. \cite{b2c,b2e}; however, then   the corresponding equations of motion are   much  more complicated and  only for constant electromagnetic field  can be more tractable.
 Namely for constant electromagnetic field and  for the  special choice $k=q/m$,  i.e. when the anomalous magnetic moment vanishes (gyromagnetic ratio equals two), e.g. for the Dirac particle,   the TD approach  simplifies significantly and  coincides with   our results.
 In fact, let us consider the constant electromagnetic field, then under  condition  \eqref{ec} we obtain that $m=M=constant$,  $P^\alpha =m\dot x^\alpha$ and 
 \be
 \label{econ}
 m\overset{..}{x}^\alpha=q{F^\alpha}_\beta\frac{d x^\beta}{d \tau }, \quad  \dot S^{\alpha\beta}=\frac{q}{m}({F^\alpha}_{\gamma}S^{\gamma\beta}-S^{\alpha\gamma}{F_{\gamma}}^{\beta}).
 	\ee
 Now,  taking $V^\alpha(0)=\dot x^\alpha(0)$ using  the condition \eqref{ec} and the first equation in \eqref{econ}  we obtain that $V^\alpha=\dot x^\alpha$; in consequence, $S^{\alpha\beta}V_\beta=S^{\alpha\beta}P_\beta$. 
 Summarizing, in this case,   we get exactly the same model as  the one obtained by means of the TD SSC  (cf. eqs. (4.6-10) in Ref. \cite{b2c}),  i.e. Thomas precession for a particle of mass $M$.  In other words, in this case the  TD condition implies also  that the  inertial part together with  the dynamical part of the momentum  gives zero. 
 Moreover, the dynamics of the centroid  $x^\alpha$ is the  same, for any $V^\alpha(0)$ (and  the same as for the $TD$ condition). 
 For other values of  the parameter $k$  such a coincidence does not hold; then, however, the situations  becomes more complicated and the full analysis is quite challenging.    
\par 
In summary, for  the  SSC    \eqref{ssc} with  \eqref{ec}  and for the factor $k$ given by eq.  \eqref{e21} there is a parametrization, $\tilde \tau $ given by  \eqref{e2},  such that $k$ and $\tilde m$   become constant.  Thus, the dynamics obtained   coincides exactly with  the one     resulting from 	symplectic structure \eqref{ep1},  \eqref{ep2} and the effective  Hamiltonian \eqref{eh}. Finally, for  the Minkowski spacetime and constant or  slowly varying electromagnetic field we  can neglect the last term in eq. \eqref{e17}. Then $x^\alpha$ satisfies the Lorentz force equation (with constant mass) while $S^{\alpha\beta}$  equation \eqref{e13}, so taking $k=q/m$ we recover the  Bargmann, Michel, and Telegdi equation (with gyromagnetic ratio equals $2$).
\par 
Finally,  let us recall that any Killing vector field (and preserving the electromagnetic field: $\mL_XF_{\alpha\beta}=0$)  yields a constant of motion of the  MPDS equation, see  e.g. \cite{b2e,b12a,b12b}.  Since we have shown that the Hamiltonian structure  \eqref{ep1}, \eqref{ep2} and \eqref{eh}  can be embedded into the  MPDS equations, 
 we immediately obtain the same relations  for  the discussed Hamiltonian formalism; this fact  was shown  by direct calculations  in   Refs. \cite{b10d,b10e}.   
 \section{Final discussion and outlook} 
 \label{s5}
 To summarize, in this work we have investigated  the motion of the spinning body  (in the pole-dipole approximation)  in the external gravitational and electromagnetic fields.   We showed that for the SSC with the  vector field   satisfying  eq. \eqref{ec}  the momentum and velocity are parallel in the presence of  electromagnetic  backgrounds in the curved spacetime.   In consequence,  the MPDS equations reduce to eqs. \eqref{e12} and \eqref{e17} with the   spin condition   given at initial time only (if necessary the  vector field  can be reconstructed by means of \eqref{ec}). In this way we obtain the whole family of the charged spinning particle models, defined by the equation of state \eqref{e18} and \eqref{e18b}, which exhibit  a  simple relation between momentum and velocity.  However, unlike the  OKS condition for gravity alone, the mass is not constant, see \eqref{e16}.  Even for  the gyromagnetic coefficient given  by eq. \eqref{e21}  (yielding the  equation    of state for the classical counterpart of the  Dirac particle) the mass dynamics remains  still nontrivial. However, in this case, by  making an appropriate   reparametrization we obtain the MPDS equations with constant mass and  gyromagnetic factor.   As a result, the model fits into the effective approach defined by  the Hamiltonian   formalism \eqref{ep1}, \eqref{ep2} and \eqref{eh}. In this way  the new
 (non-affine) parametrization, cf. \eqref{e3},   corresponds to an additional constraint in the Hamiltonian formalism (for a wider discussion of this constraint and its   time-delay consequences  see \cite{b10b}). Moreover, since   the symmetries of metric and electromagnetic  fields yield  constants  of motion   for  the  MPDS equations,  we  immediately obtain  the same conclusions  for the discussed  effective theory.
 \par
 The above results indicate that the SSC defined by the field \eqref{ec}   deserves some attention.  However, as we noted in Sec.  \ref{s2} the motion of the centroid  under this condition may  have a    different character than  for the ordinary SSC.  The  preliminary analysis     suggests that this  fact can be related to acceleration of the observer. However, in order to fully clarify this issue and the  physical meaning of the presented condition,   a more  complete investigations need to be carried out in the future, including    the so-called bobbing  motions \cite{bb1}, the range of applicability, the consequences of the  electric dipole moment    and, finally, comparison   with others conditions  (in analogy to Refs.  \cite{b3mmmm,b4a,b5a}).             
 \par
 At the end let us note that the  results obtained can be  also  a starting point for other directions of  investigation. Let us point out a  few of  them. First,  on the basis of   Refs. \cite{b8a,b8b}, it seems that  the dynamics of the  spinning particle in  the  pp-waves  metrics and corresponding (via double copy conjecture \cite{b7a}-\cite{b7e})   electromagnetic fields in the flat spacetime    can be directly related. This, in turn, would be another  example  of the  gravito-electromagnetic analogy \cite{b6a,b6b}.  In  particular,  we can analyse the motion of the  spinning particle in the  plane gravitational waves. This problem, under the  TD condition,  have been attacked in Refs. \cite{b13a,b13b};  then, however,  the equations   are quite complicated and only some particular solutions were obtained.   Applying the OKS condition with an arbitrary initial vector    we can simplify equations and   simultaneously  generalize the results of Ref.   \cite{b9}; moreover,  for the  plane  waves   considered  in \cite{b8b}, it should be  possible to find the dynamics   explicitly.     This is closely  related to the constants of motion. It turns out that  for  the spinless particles  and some special backgrounds  we can construct such constants also from proper conformal vector fields \cite{b14b}.  Thus, the question arises  whether this procedure (more generally,  the so-called nonlocal integrals of motion) can be extended to  the spinning particles. Next,   it  was shown in the work  \cite{b15a}  that the  MPDS can be obtained  as dimensional reduction of the gravitational system  in a five-dimensional  Kaluza-Klein background under the FMP condition. The question is what happens if we use  the OKS condition instead, i.e.   whether we obtain  eq. \eqref{ec} for the  reduced system.
 Finally, it would be interesting to extend the results  to the  spacetimes with non-zero torsion \cite{b15b} or  apply to  
  deflections     and  lensing  of the  massive particles \cite{b15c}. 
\vspace{1cm}
\newline 
{\bf Acknowledgment}
\par 
The  author would like to thank    Piotr  Kosi\'nski for reading manuscript and  useful comments as well as Alexei  Deriglazov for valuable e-discussion and references. 
Special thanks are due to anonymous Referees for valuable comments and suggestions,  which allowed to significantly  improve the paper.    

  \end{document}